# Opportunities for TeV Laser Acceleration

M. Kando, H. Kiriyama, J. K. Koga, S. Bulanov, A. W. Chao*, T. Esirkepov, R. Hajima, and T. Tajima

Kansai Photon Science Institute, Japan Atomic Energy Agency, Kyoto 619-0215, Japan

and

*Stanford Linear Accelerator Center, Stanford University, Stanford, California 94309, USA

**Abstract**

A set of ballpark parameters for laser, plasma, and accelerator technologies that define for electron energies reaching as high as TeV are identified. These ballpark parameters are carved out from the fundamental scaling laws that govern laser acceleration, theoretically suggested and experimentally explored over a wide range in the recent years. In the density regime on the order of $10^{16}$ cm$^{-3}$, the appropriate laser technology, we find, matches well with that of a highly efficient high fluence LD driven Yb ceramic laser. Further, the collective acceleration technique applies to compactify the beam stoppage stage by adopting the beam-plasma wave deceleration, which contributes to significantly enhance the stopping power and energy recovery capability of the beam. Thus we find the confluence of the needed laser acceleration parameters dictated by these scaling laws and the emerging laser technology. This may herald a new technology in the ultrahigh energy frontier.

## 1. Introduction

Since the laser based particle acceleration was conceived [1], energies of laser accelerated electrons have increased with the advance of laser technologies and better control and understanding of the experiments [2-12]. Leemans et al. reported one GeV laser acceleration of electrons using 40 TW, 40 fs laser pulses with a 3 cm plasma channel [2]. With these and other past experiments it is now evident that the laser accelerated electron energies scale as predicted by Tajima and Dawson [1]. The laser accelerated electron energy scales inversely proportional to the plasma electron density $n_e$. An experimental data summary is shown in Fig. 1. This figure shows this energy gain renormalized appropriately by the laser intensity (see below about this normalization) as a function of the plasma density $n_e$: Unmistakably, the energy gain rises continually and linearly, as the electron density falls. This presents us opportunities to consider a design of experiments toward 10 GeV, 100 GeV, and 1 TeV energies based on laser wakefield acceleration, following the simple and yet robust scaling law. In particular we can identify ballpark parameters required for TeV electron and positron acceleration. We find the required laser parameters match well the present or near-term laser technologies: our design principles encounter opportune novel techniques to lower required laser parameters.

## 2. Scaling laws

Summarized here are the scaling laws of laser wakefield acceleration that have been theoretically presented and experimentally observed in the past works [13][14] . If we take the scaling law according to Esarey and Sprangle [13], the energy gain of electrons per plasma stage, $\Delta E$, is approximately expressed as

$$\Delta E \approx 2m_0 c^2 a_0^2 \gamma_{ph}^2 = 2m_0 c^2 a_0^2 \left( \frac{n_{cr}}{n_e} \right), \quad (1)$$

where $m_0$ is the electron rest mass, $c$ is the speed of light, $a_0 = eE_0/m_0\omega_0 c$ is the normalized vector potential of the pump laser with the electric field of $E_0$ and the frequency of $\omega_0$, $\gamma_{ph} = \left(1 - v_{ph}^2/c^2\right)^{-1/2}$ and $v_{ph}$ is the phase velocity of the wakefield, $n_{cr}$ and $n_e$ are the critical density and the plasma density, respectively. This is based on a simple and yet robust theory of nonlinear wakefield that is one-dimensional. The reason for this may be seen from a simple argument as follows: the original energy gain theory of Tajima and Dawson gives $\Delta E = mc^2 \gamma_{ph}^2$. If electrons that participate in making up the accelerating longitudinal fields have gained their momentum (primarily in the longitudinal direction) by a factor of $a_0^2$, their mass $m$ takes the value of $m_0 a_0^2$. Obviously, in more realistic cases three-dimensional phenomena become important particularly when we try to preserve a good quality wakefield structure over many oscillation periods. On the other hand, in a relatively strong drive, the wave assumes a steep profile and thus once again nearly one-dimensional physics might become important in the immediate vicinity of this sharp gradient. In cases that we experienced in the past works such is not out of the question. In fact Koga et al.'s simulation [15] saw a steep wave gradient and much flattened wave front even though his laser pulse was relatively narrow (see Figure 2): The frontal part of the wave is appropriate for accelerating positrons [16] (or other positively charged particles), while the rear part for electrons (not all the parameters in Ref. [15] scale with what we suggest here). We believe that we still need much research to hone the parameters and fashion to reach such developments. For the sake of proof-of-principle experiments perfectly ideal realization may have to be forgone in favor of realizable ballpark parameters to gain a glimpse into really high energy regimes.

The acceleration length is limited by the dephasing length or the pump depletion length. The dephasing and pump depletion lengths are given by [13]

$$L_d = \frac{2}{\pi} \lambda_p a_0^2 \left( \frac{n_{cr}}{n_e} \right), \tag{4}$$

$$L_p = \frac{1}{3\pi} \lambda_p a_0 \left( \frac{n_{cr}}{n_e} \right), \tag{5}$$

where $\lambda_p = \sqrt{\pi / r_e n_e}$ is the linear and nonrelativistic plasma wavelength and $r_e$ is the classical electron radius. Here $\lambda_p$ is defined without the relativistic mass effect. The good condition for laser wakefield excitation is realized when the laser pulse length $c\tau$ matches this plasma wavelength, *i.e.*

$$c\tau \approx \lambda_p / 2. \tag{6}$$

The peak power of the laser $P$ is $P = I\pi w_0^2$, where $I$ is the peak irradiance of the laser. The total laser energy $E_L = P\tau$ necessary in the first case (quasi-one dimensional case) is expressed as

$$E_L[\text{J}] = \left( \frac{a_0}{0.86 \lambda_0[\mu m]} \right)^2 \times \frac{\pi (w_0[\mu m])^2}{100} \tau[\text{ps}]. \tag{7}$$

In cases when three dimensional effects are important (e. g. $w_0 \leq c/\omega_p$) the energy obtained by laser acceleration may become slightly more complicated. Consider the case when the laser pulse is intense enough to make a cavity behind the laser pulse, i.e. $a_0 > e\varphi/(mc^2)$, where $\varphi \approx 4\pi n_e e w_0^2$ is the electrostatic potential of the wake. According to a study [17], in this case

$$\Delta E \approx \frac{\pi^2 m_0 c^2 w_0^4}{8 \lambda_0^2 \lambda_p^2}, \tag{8}$$

where the laser spot size $w_0$ is related to

$$w_0 \approx \frac{1}{\pi}\gamma_{ph}\lambda_0 \qquad (9)$$

and the cavity longitudinal size is of the order of the transverse size, $w_0$. In this tightly focused case, optical guiding is required to extend the acceleration length. According to Ref. [18], the matched spot size $w_M$ in the capillary, which has a radial plasma density profile $n_e(r) = n_0 + \Delta n_e(r/R)^2$, is

$$w_M = \left(\frac{R^2}{\pi r_e \Delta n_e}\right)^{1/4}, \qquad (10)$$

where $R$ is the radius of the capillary wall. If we set $w_0 = c/\omega_p = \lambda_p/2\pi$ in order to avoid self-focusing or filamentation of the pump pulse, the accelerated energy is scaled as $\Delta E \propto 1/n_e$. In any case as shown in Fig.1 the energy of electrons scales as inversely proportional to the plasma density. This is also observed in Ref. [19]. These tendencies are in agreement with Eq. (1) and also Eq. (2).

## 3. Ballpark parameters of laser electron accelerator toward TeV

We point out that the scaling law dictates some two orders of magnitude density reduction from most current experimental parameters in order to carry out experiments in the range toward energies of TeV in a single stage, with the typical density at $10^{16}$ cm$^{-3}$. This in turn allows us to extend the laser pulse length by an order of magnitude, to typically ps, instead of tens of fs. The preferred laser technology of recent laser acceleration experiments has been that of Ti:sapphire because of its large frequency bandwidth. In spite of its superiority in its bandwidth, it has some problems such as relatively poor quantum efficiency. The choice parameters we present here allow us to introduce a different laser technology that matches better for the purpose of higher energy acceleration toward TeV. The technology we suggest is that of a laser diode-pumped Yb ceramic laser. This laser operates well in the ballpark of ps laser pulse and has high quantum efficiency and large fluence, which are important ingredients for high energy acceleration. As we find, the required laser acceleration scaling and this laser technology match well.

Here we take a few typical numerical examples at various initial laser intensities as listed in Table 1. We range the laser intensity, or the normalized vector potential $a_0$ that is related to the intensity. If we take $a_0 = 10$, $n_e = 5.7 \times 10^{16}$ cm$^{-3}$, $\lambda_0 = 1$ μm we reach 1 TeV energy in a single stage using Eq. (1). We have chosen the spot size of 32 μm which is ~1/4 of the linear nonrelativistic plasma wavelength to relax the laser energy requirement. We know this choice does not match one-dimensional scaling laws because the spot size is smaller than this plasma wavelength. However, in the nonlinear regime of the wakefield, a single cycle of the wake is used for acceleration, and the actual excited structure might be similar to that predicted by one-dimensional theory [13]. We point out that there has been no detailed simulation work in this ballpark because of the huge computational resources required. Clearly this is one of the very important future research topics to optimize the acceleration conditions in this regime. Under the current choice of ballpark the laser parameters are 0.5 kJ, 2.2 PW, and 0.23 ps. This is the case studied in Table I (a) case I. The acceleration length is ~3 m. In this choice the required laser pulse may strain the existing laser technology, as we shall discuss below. To ameliorate such a situation, the introduction of the nonuniform plasma density profile with a density initially lower than the value taken here might bring in some room to maneuver: the laser pulse compression may take place through the nonlinear interaction with the plasma [20] to fit more adequately and gradually increase the density to the value considered here.

If we decrease the laser intensity to take the value of $a_0$=3.2(Table I (a)case II), the optimum density is 5.7x10$^{15}$cm$^{-3}$. In the following we keep the ratio of the spot size to the plasma wavelength the same as that in case I. The required laser parameters are 1.6 kJ, 2.2 PW, and 0.74 ps. The acceleration length is 29 m. Further reduction of laser intensity results in longer acceleration length and larger laser energy as seen in case III in Table 1 (a). Also listed in Table 1 (b) is a study of three dimensional cases.

**4. Yb laser technologies**

These ballpark parameters for the accelerator lead to a suggestion of a laser that is robust in the pulse range of ps. As we indicated above, the acceleration toward the energies of TeV, our choice of plasma density is on the order of 10$^{16}$ cm$^{-3}$, which dictates the optimal laser pulse length on the order of O(1) ps. This is as opposed to the currently popular tens of fs pulses that are typically good for GeV or less energies driven by Ti:sapphire lasers. This implies an important point that we are now emancipated from the fs laser demand and look for a different set of technologies that allow some broader ranges of tolerance and different requirements such as higher fluence, higher efficiency, higher energy, etc. We suggest that a technology that globally meets with the need if we adopt the laser-diode (LD) pumped ytterbium (Yb)-doped ceramic laser (with a more radically different option on the horizon).

The Yb laser is a good compromise between the high energy Nd:glass lasers and the ultrabroadband Ti:sapphire lasers. Yb has a few times lower quantum defect (<10%) compared with Nd:glass. The broad absorption spectra of Yb match the emission bands of high-power InGaAs LD, which ensure approximately ten times higher efficiency compared to Ti:sapphire lasers. The fluorescence lifetime of Yb is about a few hundred times longer than that of Ti (msec vs. μsec), allowing for an enhancement of energy storage.

The recent progress of Yb ceramic laser [21] allows us to look for a large aperture thin disk design. The thin disk laser concept is a laser design for a LD pumped solid-state laser which allows the realization of lasers with high output energy, high efficiency and good beam quality, simultaneously [22]. Especially, the generation and amplification of ultra-short pulses are possible with huge laser energy and high efficiency [22]. We note that if we strain some of the laser parameters such as the pulse length (to make it shorter), the acceleration length is greatly shortened and can contribute to even more compact accelerators. On the other hand, such a choice will severely test the Yb laser technology to stomach the broader bandwidth operation. More recently, Yb-doped ceramics laser materials with an almost perfect pore-free structure was realized by advanced ceramics processing [21]. Ceramics as laser material have several remarkable advantages over single crystals. For example, samples with high doping concentration and of large sizes can be more easily fabricated, whereas this is usually difficult for crystals. The cost of ceramic laser materials can be potentially much lower than their single crystal counterparts because of their faster fabrication process and possibility of mass production. Moreover, the rigid bonding of multiple samples is easy with these materials; thus the design flexibility for novel laser devices is greatly enhanced (In fact we have developed some techniques to do so).

The technique of chirped-pulse amplification (CPA) and recompression may enable Yb-doped ceramic laser systems to provide pulses around 1 picosecond. Pulse duration in a CPA system is determined primarily by the laser output spectrum. If we introduce mixed-ceramic amplifiers, the effective gain spectral bandwidth should be increased and hence the reduction in pulse duration to well below 1 picosecond should be possible [23].

The properties of the LD directly pumped Yb-doped ceramic laser open the way to a completely new class of compact, robust, and cost effective short pulsed laser systems with highest output energies, highest efficiency, best beam quality, and high repetition rate.

In addition to the LD pumped Yb-doped ceramic laser technology, many have suggested alternatives such as fiber lasers, diode pumped alkaline lasers, etc. It is exciting to have many alternative suggestions that are now emerging for candidates of high power high fluence lasers that may fit the specifications for ultrahigh energy laser accelerators.

## 5. Compact decelerator in the TeV range

In order for a high energy accelerator system to be much more compact than following the conventional designs, it is necessary not only to make the accelerator compact, but also to make the beam dump section to be compact. For this purpose, we introduce the concept of a passive plasma decelerator at the end of the use of the high energy beam by immersing the beams to be decelerated into an appropriately designed plasma that is surrounded by a waveguide structure that has the right impedance and coupling to the oscillating plasma and extracts its electric energy.（For a linear collider, a critical problem has been its extensive power consumption. Once a TeV beam is established, luminosity is basically proportional to the power consumption. One way to ameliorate this problem is to convert the energy of one pulse of the spent TeV beam into microwaves, and utilize these microwaves to accelerate the next beam pulse）.

As one very first step of the above idea, it is necessary to decelerate a TeV-beam by some mechanism, and then convert the beam energy into microwaves. Here we consider the possibility of decelerating the TeV-beam by a plasma. The beam deposits its energy as a plasma wave of a certain plasma frequency. A passive microwave structure next to the beam passage then couples with the plasma oscillation. If the plasma frequency coincides with the microwave frequency of the structure, a microwave is excited by the oscillating plasma. As compared with this, the conventional approach of the beam decelerator is the beam dump. This is to use the single particle interaction between the high energy charged particles (such as electrons) and solid or liquid targets. Because of the single particle interaction in high energy regimes, its stopping length becomes large. In addition, the beam radiates copious radiation in the surroundings during the stopping as well as the stopping materials become radioactive in the end. In the design of the International Linear Collider, a water dump and a noble gas dump are under consideration, in which the electrons lose their energy via bremsstrahlung and the following e$^-$-e$^+$ pair creations. Our suggested method radically departs from this single particle interaction approach of the past. We resort, just as in the accelerating stage we are proposing, to use the collective electromagnetic force arising in the plasma. Just like in the acceleration, the collective force in the deceleration can be large. As Chen et al. have suggested [24] using an electron beam to drive wakefields in a plasma for the purpose of acceleration, the decelerating field is similar just reversing the accelerating process in terms of energy flow. In fact both the accelerating and the decelerating processes have been demonstrated in the recent SLAC experiment [25]. Furthermore since they are of collective nature, they are well organized and amenable to direct energy recovery into electricity. However, unlike the conventional energy recovery linac (ERL) utilizing superconducting technology [26], we recognize that it is no longer possible to bend TeV electron (positrons) orbits back to the original superconducting duct that accelerated the beam. Here the decelerator is in the down stream. The decelerating plasma is immersed in and coupled to a high-Q waveguide structure with appropriate impedance. Given a plasma of density $n_e$, its plasma frequency is given by $\omega_p = \sqrt{4\pi n_e e^2/m_0}$. Here it may be assumed that the entering gas such as Helium have the electron density of $n_e$ and its ionization is incurred by the entering intense electron beam each time. Given the TeV-beam, its decelerating gradient in the plasma is $G_{dec}$, which is on the order of the Tajima-Dawson field of

$$G_{dec} \sim m_0 \omega_p c/e. \qquad (11)$$

The plasma oscillation will last for time $\tau_p$, which is the inverse of the Landau damping decrement $\gamma_L$ before extraction of energy of the excited waveguide mode. A coupling window between the microwave (more accurately THz waves) structure and the plasma medium allows the plasma to fill the structure with microwave. (If we assume a plasma density of $10^{14}$ cm$^{-3}$, the exciting rf wave becomes 90 GHz, which is a familiar frequency as W-band and many industrial components are available.) A traveling wave rf structure is to be designed that couples into this plasma rf source. The input plasma rf power will be fed into the structure with critical coupling so that there is no reflected power. The higher the plasma temperature, the smaller the $\tau_p$ is due to the Landau damping. We dictate that the energy extraction time from the waveguide shorter that this damping time of $\tau_p$. The details are provided by Chao and Tajima [27].

**6. Discussion and conclusions**

Even though what we suggest here is merely a crude outline for a proof-of-principle experimentation, it is possible to gain an uncanny amount of knowledge by testing such regimes of laser acceleration. We have already identified several key technologies that support the above systems. It is important to investigate the validity of the parameter scaling laws further into 100 GeV, and even 1 TeV. The stability of the laser pulse over a distance of such magnitude never experienced before is a critical question to be examined by experiments and perhaps also by a really huge scale simulation. The propagated laser may be tinted with spectral broadening and pulse lengthening (or shortening). It is also of interest to investigate if there are ways to further improve the way to excite wakefields and to explore optimal parameter ballpark and method of experimentation. If the acceleration conditions that impose requirements on laser and the laser technology that can be delivered clash with each other, are there ways to navigate the best course of action to compromise? Which parameter or parameters can be bent and which cannot? Can we introduce ways novel or otherwise to tolerate errors and to invite robustness in certain regimes? Which parameter regimes are obstinate to even a small amount of change from the optimal values?

The focal length of the lens to match the spot size should be taken into account when we discuss the size of the whole system. The spot size focused with a focal length of $f$ is expressed as $w_0 \approx (f/D)\lambda_0$, where $D$ is the beam size at focusing optics. The damage threshold limits the size of $D$. For example, usual multilayer coating has a damage threshold of ~ 5 J/cm$^2$. This gives D ~16 cm for $E_L = 1$ kJ for the quasi-one dimensional case and D ~ 71 cm for $E_L = 20$ kJ for 3D case. Thus the focal length is ~16 m when $w_0 = 100$ μm (quasi-1D) and ~ 320 m when

$w_0$ =450 μm (3D).

We can envisage experiments that reach energies of 10 GeV and 100 GeV on our way toward still higher energies. In any of these cases the strategy to pick plasma density and other subsequent parameters may vary. In an overall tendency we are allowed to take higher density, shorter laser pulse, shorter acceleration length, and less laser energy in the cases lower than TeV, but with approximately similar power as for the TeV acceleration, though clearly there are many easier experiments with other aspects emphasized for specific convenience or technological reasons. For example, experiments for 10 GeV may not even need any change of laser technology, continuing the usage of the Ti:Sapphire and simply extending the acceleration length from O(1) cm to O(10) cm. For 100 GeV we may have to reduce the plasma density from the currently common $10^{18-19}$ cm$^{-3}$ to the range on the order of $10^{17}$ cm$^{-3}$. Even before we reach energies of TeV, at these energies of 10 GeV and 100 GeV with the laser acceleration present us unprecedented opportunities of new frontiers in many directions. For example, because the laser-driven electron beam pulse is ultrashort in fs or even in as regimes, we can make bright coherent X-rays out of 10 GeV laser accelerated electron beam that can be generated within a foot size accelerator as opposed to 100 m size installations. Such a table-top bright alternative to a synchrotron radiation source may become available at each laboratory that needs one. With beams on the order of 100 GeV, we may entertain a compact and yet serious experiments and facilities that may not be imaginable with the conventional technologies. For example, these compact 100 GeV electrons and a sliced out portion of the laser may be allowed to collide to produce bright, collimated, mono-energetic and energy-tunable, Compton-backscattered gamma rays. Such gamma ray beams have never been produced before. It is hardly within our grasp what discovery awaits us. In addition with even much lower energies of electrons we can explore nuclear physics and energy science relevant issues with Compton backscattered gamma rays [28]. We also expect the proposed compact plasma decelerator may be of use in not only a future collider but also in lower energy devices.

In conclusion we have identified the opportune confluence of the conditions for laser acceleration toward 1 TeV and the novel technology of a high energy, high power laser of Yb base and the subsequent scientific opportunities that this confluence opens up. We also note that it is crucial to reduce the size of the beam dump or more elegantly to introduce an energy-recovering compact decelerator at its rear end. A first such a suggestion is made. We face many research tasks that need to be explored to realize this opportunity.

We appreciate the camaraderie of our colleagues Dr. H. Kotaki, Dr. Y. Fukuda, Dr. Y. Kato, Dr. T. Kimura, Dr. S. Kawanishi, Dr. K. Nakajima, Dr. H. Daido, Mr. Y. Hayashi, Dr. I. Daito, Mr. T. Homma, Mr. S. Kondo, and Mr. S. Kanazawa among many who contributed to this.**References**

[1] T. Tajima and J. M. Dawson, Phys. Rev. Lett., **43**, 267, (1979).

[2] W. P. Leemans, et al., Nature Physics, **2**, 696 (2006).

[3] E. Miura, et al., Appl. Phys. Lett. **86**, 251501 (2005).

[4] S. P. D. Mangles, et al., Nature, **431**, 535 (2004).

[5] C. G. R. Geddes, et al., ibid, 538 (2004).

[6] J. Faure, et al., ibid, 541 (2004).

[7] A. Yamazak,i et al., Phys. Plasmas,**12**, 093101 (2004).

[8] T. Hosokai, et al., Phys. Rev. E, **73**, 036407 (2006).

[9] C.-T. Hsieh, et al., Phys. Rev. Lett. , **96**, 095001 (2006).

[10] B. Hidding, et al., Phys. Rev. Lett., **96**, 105004 (2006).

[11] M. Mori, et al., Phys. Lett. A, **356**, 146 (2006).

[12] S. P. D. Mangles et al., Phys. Rev. Lett., **96**, 215001 (2006).

[13] E. Esarey and P. Sprangle, IEEE Trans. on Plasma Science, **24**, 252 (1996).

[14] W. Lu, et al., Phys. Rev. Lett., **96**, 165002 (2006).

[15] J. Koga, K. Nakajima, and K. Nakagawa, in *Superstrong Fields in Plasma*, ed. M. Lontano et al. (AIP, New York, 2001), p.126 (2002).

[16] T. Zh. Esirkepov, et al., Phys. Rev. Lett., **96,** 014803 (2006).

[17] S. V. Bulanov, Plasma Phys. Control. Fusion **48**, B29 (2006).

[18] N. A. Bobrova, et al., Phys. Rev. E **65**, 016407 (2002).

[19] A. Maksimchuk et al., Appl. Phys. B **89**, 201 (2007).

[20] J. Faure, et al., Phys. Rev. Lett., **95**, 205003 (2005).

[21] A. Ikesue, et al., J. Am. Ceram. Soc., **89**, 1936 (2006).

[22] A. Giesen, et al., Appl. Phys. B **58,** 365 (1994).

[23] I. N. Ross, et al., Appl. Opt. **36**, 9348 (1997).

[24] P. Chen, et al., Phys. Rev. Lett., **54**, 693-696 (1985).11

**Figure and table captions**

Figure 1. Electron energy vs. plasma density observed in experiments [2-12]. The solid line shows the fitted curve of $\Delta E/a_0^2/mc^2 = 5.5 \times 10^{21}/n_e$ gleaned and $\chi^2$-matched from all these experimental data, and the broken line shows the theoretical scaling ($\Delta E/a_0^2/mc^2 = 1.7 \times 10^{21}/n_e$).

Figure 2. Two dimensional particle-in-cell simulation results reproduced from ref [15] for a 19fs 100 TW laser pulse with a wavelength of 0.8μm propagating in an uniform plasma of density $5.3 \times 10^{19}$ cm$^{-3}$. After propagating $340c/\omega_{pe}$, (a) the laser electric field with a peak amplitude of $a_0 \sim 13$, (b) electron density with a peak of $1.3 \times 10^{21}$cm$^{-3}$, and (c) electric field with a peak of 6.5 TeV/m are shown. One can see that the structure is nearly one-dimensional.

Table 1. Example parameters for TeV proof-of-principle laser acceleration of electrons and positrons.

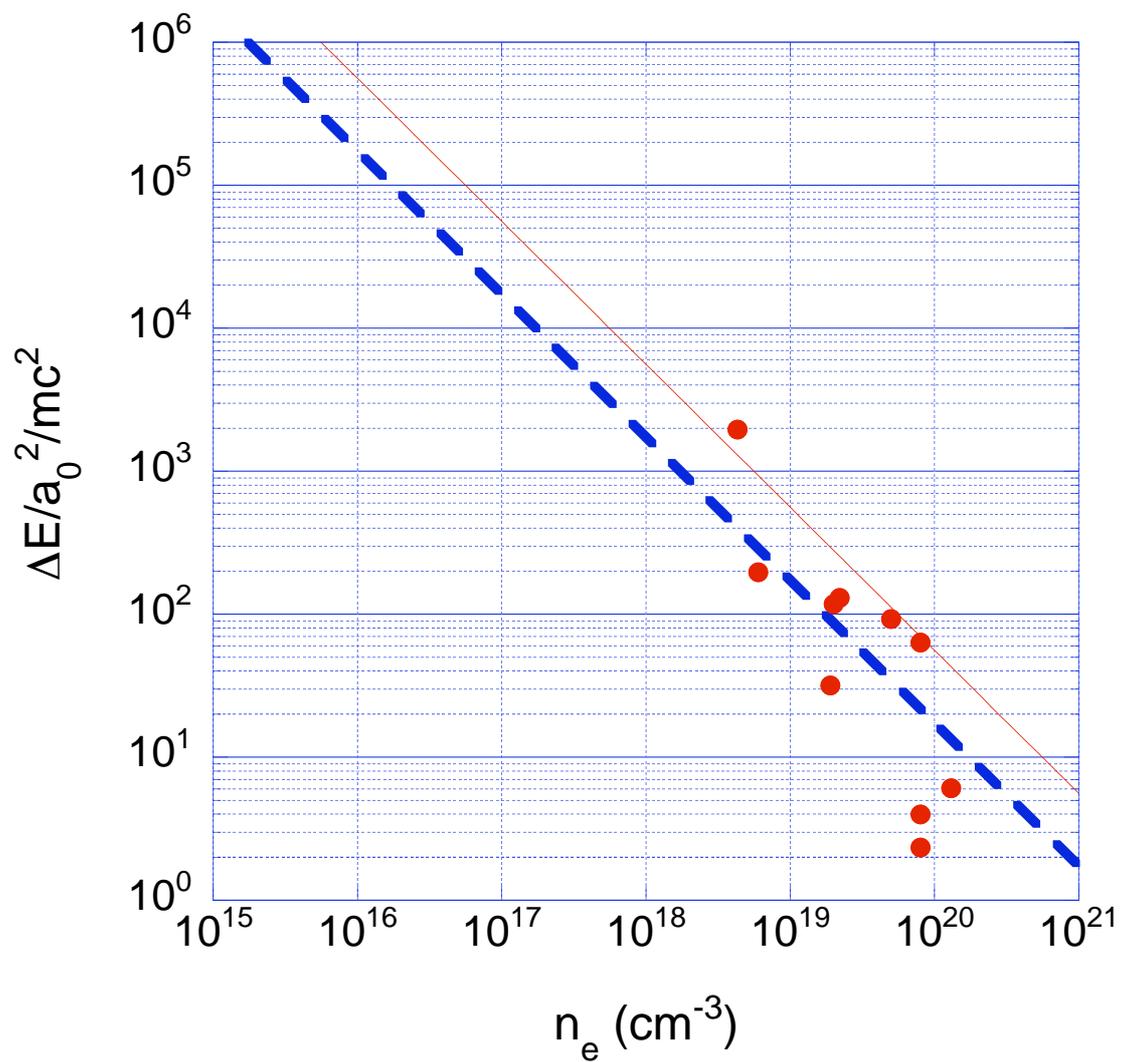

Figure. 1

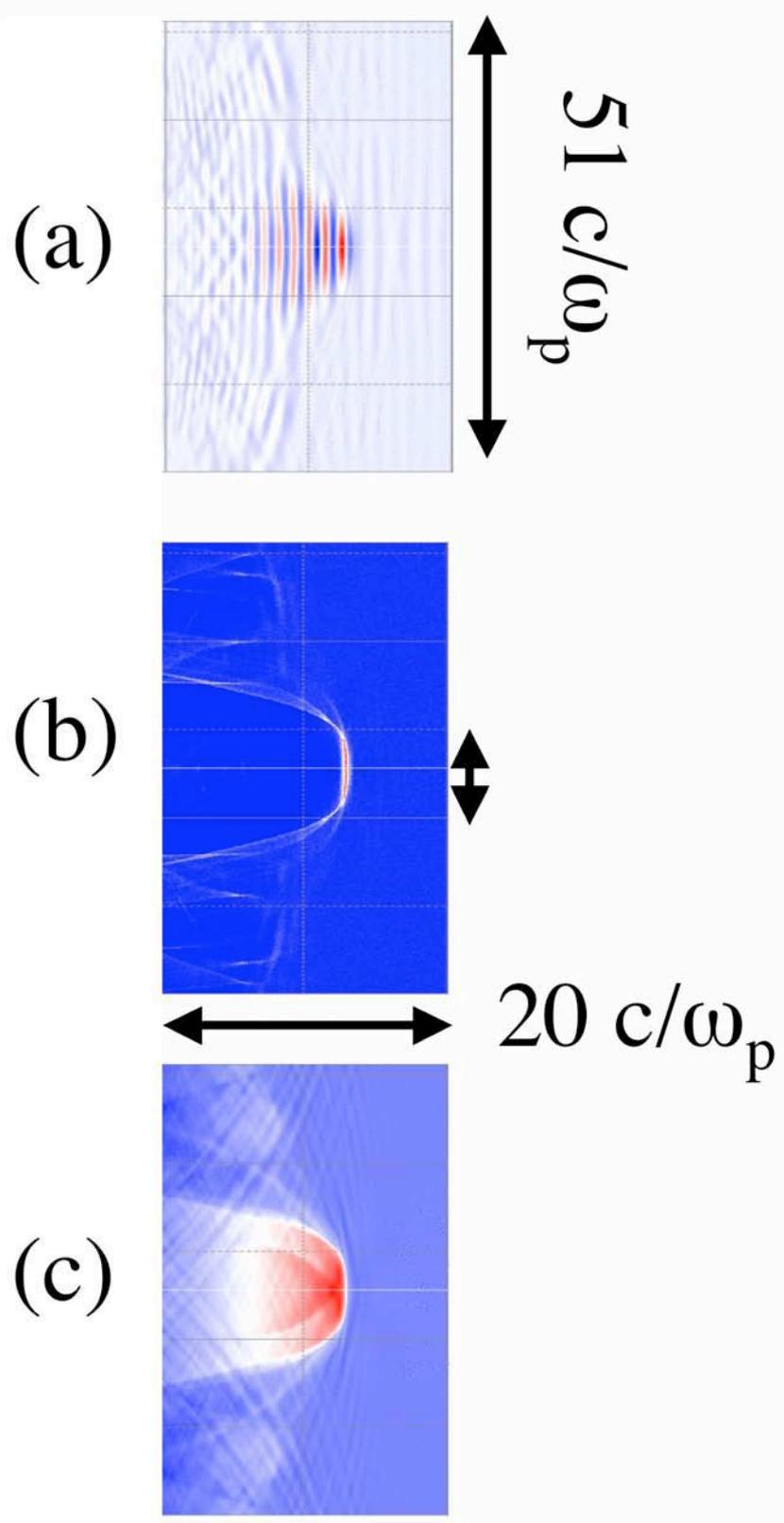

Figure 2

Table 1 (a).

|  |  | case I | case II | case III |
|---|---|---|---|---|
| $a_0$ |  | 10 | 3.2 | 1 |
| energy gain | GeV | 1000 | 1000 | 1000 |
| plasma density | cm$^{-3}$ | 5.7x10$^{16}$ | 5.7x10$^{15}$ | 5.7x10$^{14}$ |
| acceleration length | m | 2.9 | 29 | 290 |
| spot radius | μm | 32 | 100 | 320 |
| peak power | PW | 2.2 | 2.2 | 2.2 |
| pulse duration | ps | 0.23 | 0.74 | 2.3 |
| laser pulse energy | kJ | 0.5 | 1.6 | 5 |

Table 1 (b).

|  |  | case IV | case V | case VI |
|---|---|---|---|---|
| $a_0$ |  | 1 | 1 | 1 |
| energy gain | GeV | 10 | 100 | 1000 |
| plasma density | cm$^{-3}$ | 7.0x10$^{16}$ | 7.0x10$^{15}$ | 7.0x10$^{14}$ |
| acceleration length | m | 1.8 | 58 | 1800 |
| capillary wall diameter | mm | 0.38 | 1.2 | 3.8 |
| spot radius | μm | 130 | 400 | 1300 |
| peak power | PW | 0.68 | 6.8 | 68 |
| pulse duration | ps | 0.21 | 0.67 | 2.1 |
| laser pulse energy | kJ | 0.14 | 1 | 20 |